\documentclass[proceedings]{stacs}
\stacsheading{2010}{477-488}{Nancy, France}
\firstpageno{477}

\usepackage{amsthm}
\usepackage{amssymb}
\usepackage{amsmath}
\usepackage{bm}
\usepackage{graphicx}
\usepackage{url}
\usepackage{verbatim}

\theoremstyle{plain}
\newtheorem{ourtheorem}{Theorem}
\newtheorem{ourlemma}{Lemma}
\newtheorem{ourcorollary}{Corollary}
\newtheorem{ourproposition}{Proposition}

\newtheorem*{lemmaE}{Lemma E}
\newtheorem*{lemmaA}{Lemma A}
\newtheorem{sublemma}{Lemma}
\numberwithin{sublemma}{ourlemma}
\newcommand{\citex}[2]{\cite[\textsc{#2}]{#1}}

\makeatletter
\newcommand{\hacklabel}[1]{\protected@edef\@currentlabel{#1}}
\makeatother

\newcommand{\D}[1]{\texttt{\textup{#1}}}
\newcommand{\basek}[2][p]{\bm{(}#2\bm{)}_{#1}}
\newcommand{\baseseven}[1]{\basek[7]{#1}}
\newcommand{\basetwo}[1]{\basek[2]{#1}}
\newcommand{\dotminus}{\mathop{\mbox{$-^{\hspace{-.5em}\cdot}\,\,$}}}

\newcommand{\removeone}{Remove_{\D1}} 
\newcommand{\appendthreesix}{Append_{\D3,\D6}} 
\newcommand{\Resolve}{Resolve}
\newcommand{\Goal}{Goal}
\newcommand{\Admissible}{Admissible}

\newcommand{\embed}[2]{\tau_{#1}(#2)}
\newcommand{\fulltrack}[1]{\embed{#1}{\mathbb{Z}}}

\newcommand{\makeset}[2]{\ensuremath{ \{ #1 \: | \: #2 \} }}
\newcommand{\makesetbig}[2]{\ensuremath{ \big\{ \: #1 \; \big| \; #2 \: \big\} }}

\renewcommand{\emptyset}{\varnothing}
\renewcommand{\epsilon}{\varepsilon}

\begin{document}

\title{On equations over sets of integers}

\author[wroclaw]{A. Je\.z}{Artur Je\.z}
\address[wroclaw]{Institute of Computer Science,
	University of Wroc{\l}aw}
\email{aje@ii.uni.wroc.pl}

\author[akatemia,turku]{A. Okhotin}{Alexander Okhotin}
\address[akatemia]{Academy of Finland}
\address[turku]{Department of Mathematics, University of Turku, Finland}
\email{alexander.okhotin@utu.fi}

\thanks{
Research supported
by the Polish Ministry of Science and Higher Education
under grants N~N206 259035 2008--2010 and N~N206 492638 2010--2012,
and by the Academy of Finland
under grant 134860.
}

\keywords{Language equations,
computability,
arithmetical hierarchy,
hyper-arithmetical hierarchy.}
\subjclass{F.4.3 (Formal languages), F.4.1 (Mathematical logic)}

\begin{abstract}
\noindent
Systems of equations with sets of integers as unknowns are considered.
It is shown that the class of sets
representable by unique solutions of equations
using the operations of union and addition $S+T=\makeset{m+n}{m \in S, \: n \in T}$
and with ultimately periodic constants
is exactly the class of hyper-arithmetical sets.
Equations using addition only
can represent every hyper-arithmetical set
under a simple encoding.
All hyper-arithmetical sets can also be represented
by equations over sets of natural numbers
equipped with union, addition and subtraction
$S \dotminus T=\makeset{m-n}{m \in S, \: n \in T, \: m \geqslant n}$.
Testing whether a given system has a solution
is $\Sigma^1_1$-complete for each model.
These results, in particular, settle the expressive power
of the most general types of language equations,
as well as equations over subsets of free groups.
\end{abstract}

\maketitle

\sloppy

\section{Introduction}

Language equations are 
equations with formal languages as unknowns.
The simplest such equations are the context-free grammars \cite{GinsburgRice},
as well as their generalization, the conjunctive grammars \cite{Conjunctive}.
Many other types of language equations
have been studied in the recent years,
see a survey by Kunc~\cite{Kunc_survey},
and most of them were found to have strong connections to computability.
In particular,
for equations with concatenation and Boolean operations
it was shown by Okhotin~\cite{DecisionProblems,EquationsUnresolved}
that the class of languages
representable by their unique (least, greatest) solutions
is exactly the class of recursive (r.e., co-r.e.) sets.
A computationally universal equation of the simplest form
was constructed by Kunc~\cite{Kunc},
who proved that the greatest solution of the equation $XL=LX$,
where $L \subseteq \{a,b\}^*$ is a finite constant language,
may be co-r.e.-complete.

A seemingly trivial case of language equations
over a \emph{unary alphabet} $\Omega=\{a\}$
has recently been studied.
Strings over such an alphabet may be regarded as natural numbers,
and languages accordingly become sets of numbers.
As established by the authors~\cite{EquationsUnary},
these equations are as powerful
as language equations over a general alphabet:
a set of natural numbers is representable by a unique solution
of a system with union and elementwise addition
if and only if it is recursive. 
Furthermore, even without the union operation
these equations remain almost as powerful~\cite{EquationsUnaryPlus}:
for every recursive set $S \subseteq \mathbb{N}$,
its encoding $\sigma(S) \subseteq \mathbb{N}$
satisfying $S=\makeset{n}{16n+13 \in \sigma(S)}$
can be represented by a unique solution
of a system using addition only, as well as ultimately periodic constants.
At the same time, as shown by Lehtinen and Okhotin~\cite{EquationsUnaryPlus2},
some recursive sets are not representable without an encoding.

Equations over sets of numbers
are, on one hand, interesting on their own
as a basic mathematical object.
On the other hand, these equations
form a very special case of language equations
with concatenation and Boolean operations,
which turned out to be as hard as the general case,
and this is essential for understanding language equations.
However, it must be noted that these cases do not exhaust
all possible language equations.
The recursive upper bound on unique solutions~\cite{DecisionProblems}
is applicable only to equations
with \emph{continuous} operations on languages,
and using the simplest non-continuous operations,
such as homomorphisms
or quotient~\cite{EquationsUniversality},
leads out of the class of recursive languages.
In particular, a quotient with regular constants
was used to represent all sets in the arithmetical hierarchy~\cite{EquationsUniversality}.

The task is to find a natural limit
of the expressive power of language equations,
which would not assume continuity of operations.
As long as operations on languages
are expressible in first-order arithmetic
(which is true for every common operation),
it is not hard to see that
unique solutions of equations with these operations
always belong to the family
of \emph{hyper-arithmetical sets}~\cite{Moschovakis,Robinson,Rogers}.
This paper shows that this obvious upper bound
is in fact reached already in the case of a unary alphabet.

To demonstrate this,
two abstract models dealing with sets of numbers
shall be introduced.
The first model are
equations over sets of natural numbers
with addition $S+T=\makeset{m+n}{m \in S, \: n \in T}$
and subtraction
$S \dotminus T=\makeset{m-n}{m \in S, \: n \in T, \: m \geqslant n}$
(corresponding to concatenation and quotient of unary languages),
as well as set-theoretic union.
The other model has sets of integers, including negative numbers, as unknowns,
and the allowed operations are addition and union.
The main result of this paper
is that unique solutions of systems of either kind
can represent every \emph{hyper-arithmetical} set of numbers.

The base of the construction is
the authors' earlier result~\cite{EquationsUnary}
on representing every recursive set
by equations over sets of natural numbers with union and addition.
In Section~\ref{section_basic_expressive_power},
this result is adapted
to the new models introduced in this paper.
The next task is representing every set in the arithmetical hierarchy,
which is achieved in Section~\ref{section_AH}
by simulating existential and universal quantifiers
over a recursive set.
These arithmetical sets
are then used in Section~\ref{section_HA} as constants
for the construction of equations
representing hyper-arithmetical sets.
Finally, the constructed equations
are encoded in Section~\ref{section_addition_only}
using equations over sets of integers
with addition only and periodic constant sets.

This result brings to mind a study by Robinson~\cite{Robinson},
who considered equations,
in which the unknowns
are functions from $\mathbb{N}$ to $\mathbb{N}$,
the only constant is the successor function
and the only operation is superposition,
and proved that a function is representable
by a unique solution of such an equation
if and only if 
it is hyper-arithmetical.
Though these equations deal with objects
different from sets of numbers,
there is one essential thing in common:
in both results, unique solutions
of equations over second-order arithmetical objects
represent hyper-arithmetical sets.

Some more related work can be mentioned.
Halpern~\cite{Halpern}
studied the decision problem
of whether a formula of Presburger arithmetic with set variables
is true for all values of these set variables,
and showed that it is $\Pi^1_1$-complete.
The equations studied in this paper
can be regarded as a small fragment of Presburger arithmetic
with set variables.

Another relevant model are languages over free groups,
which have been investigated, in particular,
by Anisimov~\cite{Anisimov}
and by d'Alessandro and Sakarovitch~\cite{dAlessandroSakarovitch}.
Equations over sets of integers
are essentially equations for languages over a monogenic free group.

An important special case of equations over sets of numbers
are \emph{expressions} and \emph{circuits}
over sets of numbers,
which are equations without iterated dependencies.
Expressions and circuits over sets of natural numbers
were studied by
McKenzie and Wagner~\cite{McKenzieWagner},
and a variant of these models
defined over sets of integers
was investigated by Travers~\cite{Travers}.

\section{Equations and their basic expressive power}
\label{section_basic_expressive_power}

The subject of this paper are systems of equations of the form
\begin{equation*}
\left\{\begin{array}{rcl}
	\varphi_1(X_1, \ldots, X_n) &=& \psi_1(X_1, \ldots, X_n) \\
	&\vdots& \\
	\varphi_m(X_1, \ldots, X_n) &=& \psi_m(X_1, \ldots, X_n)
\end{array}\right.
\end{equation*}
where $X_i \subseteq \mathbb{Z}$
are unknown sets of integers,
and the expressions $\varphi_i$ and $\psi_i$
use such operations as union, intersection, complementation,
as well as the main arithmetical operation
of elementwise addition of sets, defined as
$S+T=\makeset{m+n}{m \in S, \: n \in T}$.
Subtraction
$S-T=\makeset{m-n}{m \in S, \: n \in T}$
shall be occasionally used.
The constant sets contained in a system
sometimes will be singletons only,
sometimes any ultimately periodic constants will be allowed
(a set of integers $S \subseteq \mathbb{Z}$ is \emph{ultimately periodic}
if there exist numbers $d \geqslant 0$ and $p \geqslant 1$,
such that $n \in S$ if and only if  $n+p \in S$
for all $n$ with $|n| \geqslant d$),
and in some cases the constants
will be drawn from wider classes of sets,
such as all recursive sets.
Systems over sets of natural numbers
shall have subsets of $\mathbb{N}$
both as unknowns and as constant languages;
whenever subtraction is used in such equations,
it will be used in the form
$S \dotminus T = (S - T) \cap \mathbb{N}$.

Consider systems with a unique solution.
Every such system can be regarded as a specification of a set,
and for every type of systems there is a natural question
of what kind of sets can be represented
by unique solutions of these systems.
For equations over sets of natural numbers,
these are the recursive sets:

\begin{ourproposition}[Je\.z, Okhotin~\citex{EquationsUnary}{Thm.\,4}]
\label{recursive_set_representable_proposition}
The family of sets of natural numbers
representable by unique solutions 
of systems of equations of the form
	$\varphi_i(X_1, \ldots, X_n)=\psi_i(X_1, \ldots, X_n)$
with union, addition and singleton constants,
is exactly the family of recursive
sets.
\end{ourproposition}

Turning to the more general cases
of equations over sets of integers
and of equations over sets of natural numbers
with subtraction,
an upper bound on their expressive power
can be obtained by reformulating a given system
in the notation of first-order arithmetic.

\begin{ourlemma}
For every system of equations in variables $X_1, \ldots X_n$
using operations  expressible in first-order arithmetic
there exists an arithmetical formula $Eq(X_1, \ldots, X_n)$,
where $X_1, \ldots, X_n$ are free second-order variables,
such that $Eq(S_1, \ldots, S_n)$ is true
	if and only if 
$X_i=S_i$ is a solution of the system.
\end{ourlemma}
Constructing this formula is only a matter of reformulation.
As an example, an equation $X_i=X_j+X_k$
is represented by 
$(\forall n) \big[n \in X_i \leftrightarrow (\exists n')(\exists n'') n=n'+n''
\land n' \in X_j \land n'' \in X_k\big]$.

Now consider the following formulae of second-order arithmetic:
\begin{align*}
	\varphi(x) &= (\exists X_1) \ldots (\exists X_n)
		Eq(X_1, \ldots, X_n) \land x \in X_1 \\
	\varphi'(x) &= (\forall X_1) \ldots (\forall X_n)
		Eq(X_1, \ldots, X_n) \to x \in X_1
\end{align*}
The formula $\varphi(x)$
represents the membership of $x$
in \emph{any} solution of the system,
while $\varphi'(x)$ states
that \emph{every} solution of the system
contains $x$.
Since, by assumption, the system has a unique solution,
these two formulae are equivalent
and each of them specifies the first component
of this solution.
Furthermore, $\varphi$ and $\varphi'$
belong to the classes $\Sigma^1_1$ and $\Pi^1_1$, respectively,
and accordingly the solution
belongs to the class $\Delta^1_1=\Sigma^1_1 \cap \Pi^1_1$,
known as the class of \emph{hyper-arithmetical sets}~\cite{Moschovakis,Rogers}.

\begin{ourlemma}
For every system of equations in variables $X_1, \ldots X_n$
using operations and constants expressible in first-order arithmetic
that has a unique solution $X_i=S_i$,
the sets $S_i$ are hyper-arithmetical.
\end{ourlemma}

Though this looks like a very rough upper bound,
this paper actually establishes the converse,
that is, that every hyper-arithmetical set
is representable by a unique solution of such equations.
The result shall apply to equations of two kinds:
over sets of integers with union and addition,
and over sets of natural numbers with union, addition and subtraction.
In order to establish the properties
of both families of equations
within a single construction,
the next lemma introduces a general form of systems
that can be converted to either of the target types of systems:

\begin{ourlemma}\label{intermediate_form_to_target_equations_lemma}
Consider any system
of equations $\varphi(X_1, \ldots, X_m)=\psi(X_1, \ldots, X_m)$
and inequalities $\varphi(X_1, \ldots, X_m)\subseteq\psi(X_1, \ldots, X_m)$
over sets of natural numbers
that uses the following operations:
union;
addition of a recursive constant;
subtraction of a recursive constant;
intersection with a recursive constant.
Assume that the system has a unique solution $X_i=S_i \subseteq \mathbb{N}$.
Then there exist:
\begin{enumerate}
\item	a system of equations
	over sets of natural numbers
	in variables $X_1, \ldots, X_m, Y_1, \ldots, Y_{m'}$
	using the operations of addition, subtraction and union
	and singleton constants,
	which has a unique solution
	with $X_i=S_i$;
\item	a system of equations
	over sets of integers
	in variables $X_1, \ldots, X_m, Y_1, \ldots, Y_{m'}$
	using the operations of addition and union,
	singleton constants and the constants $\mathbb{N}$ and $-\mathbb{N}$,
	which has a unique solution
	with $X_i=S_i$.
\end{enumerate}
\end{ourlemma}

Inequalities $\varphi \subseteq \psi$
can be simulated by equations $\varphi \cup \psi=\psi$.
For equations over sets of natural numbers,
each recursive constant is represented
according to Proposition~\ref{recursive_set_representable_proposition},
and this is sufficient to implement
each addition or subtraction of a recursive constant
by a large subsystem using only singleton constants.
In order to obtain a system over sets of integers,
a straightforward adaptation of Proposition~\ref{recursive_set_representable_proposition}
is needed:

\begin{sublemma}
\label{recursive_set_representable_sublemma}
For every recursive set $S \subseteq \mathbb{N}$
there exists a system of equations over sets of integers
in variables $X_1, \ldots, X_n$
using union, addition, singleton constants
and constant $\mathbb{N}$,
such that the system has a unique solution
with $X_1=S$.
\end{sublemma}

This is essentially the system
given by Proposition~\ref{recursive_set_representable_proposition},
with additional equations
$X_i \subseteq \mathbb{N}$.

Now a difference $X \dotminus R$
for a recursive constant $R \subseteq \mathbb{N}$
shall be represented as $(X + (-R)) \cap \mathbb{N}$,
where the set $-R = \makeset{-n}{n \in R}$
is specified by taking a system for $R$
and applying the following transformation:

\begin{sublemma}[Representing sets of opposite numbers]\label{opposite_numbers_lemma}
Consider a system of equations over sets of integers,
in variables $X_1, \ldots, X_n$,
using 
union and addition,
and any constant sets,
which has a unique solution $X_i=S_i$.
Then the same system, with each constant $C \subseteq \mathbb{Z}$
replaced by the set of the opposite numbers $-C$,
has the unique solution $X_i=-S_i$.
\end{sublemma}

The last step in the proof of Lemma~\ref{intermediate_form_to_target_equations_lemma}
is eliminating intersection with recursive constants.
This is done as follows:

\begin{sublemma}[Intersection with constants]
\label{intersection_with_recursive_constants_lemma}
Let $R \subseteq \mathbb{N}$ be a recursive set.
Then there exists a system of equations
over sets of natural numbers
using union, addition and singleton constants,
which has variables $X, Y, Y', Z_1, \ldots, Z_m$,
such that the set of solutions of this system is
\begin{equation*}
	\makesetbig{
	(X=S, \;
	Y=S \cap R, \;
	Y'=S \cap \overline{R}, \;
	Z_i=S_i)}
	{ S \subseteq \mathbb{N} },
\end{equation*}
where $S_1, \ldots, S_m$ are some fixed sets.
\end{sublemma}
In plain words, the constructed system works
as if an equation $Y=X \cap R$
(and also as another equation $Y'=X \cap \overline{R}$,
which may be ignored).
This completes the transformations
needed for Lemma~\ref{intermediate_form_to_target_equations_lemma}.

The last basic element of the construction
is representing a set of integers (both positive and negative)
by first representing its positive and negative subsets
individually:

\begin{ourlemma}[Assembling positive and negative subsets]\label{assembling_negative_and_positive_lemma}
Let sets $S \cap \mathbb N$ and $(-S) \cap \mathbb N$
be representable by unique solutions
of equations over sets of integers
using union, addition, and ultimately periodic constants.
Then $S$ is representable by equations over integers using only union, addition and ultimately periodic constants.
\end{ourlemma}

\section{Representing the arithmetical hierarchy} \label{section_AH}

Each arithmetical set can be represented
by a recursive relation with a quantifier prefix,
and arithmetical sets form the \emph{arithmetical hierarchy}
based on the number of quantifier alternations in such a formula.
The bottom of the hierarchy are the recursive sets,
and every next level is comprised of two classes, $\Sigma^0_k$ or $\Pi^0_k$,
which correspond to the cases of the first quantifier's being existential or universal.
For every $k \geqslant 1$, a set is in $\Sigma^0_k$
if it can be represented as
\begin{equation*}
	\makeset{w}{\exists x_1 \forall x_2 \ldots Q_k x_k \; R(w, x_1, \ldots, x_k)}
\end{equation*}
for some recursive relation $R$, where $Q_k=\forall$ if $k$ is even and
$Q_k=\exists$ if $k$ is odd.
A set is in $\Pi^0_k$ if it admits a similar representation
with the quantifier prefix $\forall x_1 \exists x_2 \ldots Q_k x_k$.
It is easy to see that $\Pi^0_k=\makeset{L}{\overline{L} \in \Sigma^0_k}$.
The sets $\Sigma^0_1$ and $\Pi^0_1$ are the recursively enumerable sets and
their complements, respectively.
The arithmetical hierarchy is known to be strict:
$\Sigma^0_k \subset \Sigma^0_{k+1}$ and
$\Pi^0_k \subset \Pi^0_{k+1}$ for every $k \geqslant 0$.
Furthermore, for every $k \geqslant 1$ the inclusion
$\Sigma^0_k \cup \Pi^0_k \subset \Sigma^0_{k+1} \cap \Pi^0_{k+1}$
is proper, i.e., there is a gap between the $k$-th and $(k+1)$-th level.

For this paper, the definition of arithmetical sets
shall be arithmetized in base-7 notation\footnote{%
	Base $7$ is the smallest base, 
	for which the details of the constructions
	could be conveniently implemented.
}
as follows:
a set $S \subseteq \mathbb{N}$ is in $\Sigma^0_k$
if it is representable as
\begin{multline*}
	S
		=
	\{\, \baseseven{w} \mid
	\exists x_1 \in \{\D3, \D6\}^* \: \forall x_2 \in \{\D3, \D6\}^*
	\ldots
	Q_k x_k \in \{\D3, \D6\}^*
	\baseseven{\D1 x_1 \D1 y_1 \D1 \ldots x_k \D1 y_k \D1 w} \in R \},
\end{multline*}
for some recursive set $R \subseteq \mathbb{N}$,
where $\baseseven{w}$ for $w \in \{\D0, \D1, \ldots, \D6\}^*$
denotes the natural number with base-7 notation $w$.
The strings $x_i \in \{\D3, \D6\}^*$
represent \emph{binary} notation of some numbers,
where $\D3$ stands for zero and $\D6$ stands for one.
The notation $\basetwo{x}$ for $x \in \{\D3, \D6\}^*$
shall be used to denote the number represented by this encoding.
The digits $\D1$ act as separators.
Throughout this paper,
the set of base-7 digits $\{\D0, \D1, \ldots, \D6\}$
shall be denoted by $\Omega_7$.

In general, the construction
of a system of equations representing the set $S$
begins with representing $R$,
and proceeds with evaluating the quantifiers,
eliminating the prefixes
$\D1 x_1$, $\D1 x_2$, and so on until $\D1 x_k$.
In the end, all numbers $\baseseven{\D1w}$ with $\baseseven{w} \in S$
will be produced.
These manipulations can be expressed
in terms of the following three functions:
\begin{align*}
	\removeone(X)
		&=
	\makeset{\baseseven{w}}{\baseseven{\D1w} \in X},
	\\
	E(X)
		&=
	\makeset{\baseseven{\D1w}}{%
	\exists x \in \{\D3, \D6\}^*: \:
	\baseseven{x\D1w} \in X},
	\\
	A(X)
		&=
	\makeset{\baseseven{\D1w}}{%
	\forall x \in \{\D3, \D6\}^*: \:
	\baseseven{x\D1w} \in X}.
\end{align*}

The expression converting numbers of the form $\baseseven{\D1w}$
to $\baseseven{w}$
is constructed as follows:
\begin{ourlemma}[Removing leading digit $\D1$]
\label{removing_leading_digit_1_lemma}
The value of the expression 
\begin{equation}\label{removing_leading_digit_1_lemma__expression}
(X \!-\! \{1\} \cap \{0\})
\cup
\bigcup_{i \in \Omega_7 \setminus \{\D0\}} \;
\bigcup_{t \in \{0, 1\}}
\big[(X \cap \baseseven{\D1 i \Omega_7^t (\Omega_7^2)^*})
\dotminus \baseseven{\D1\D0^*}\big]
\cap \baseseven{i \Omega_7^t (\Omega_7^2)^*}
\end{equation}
on any $S \subseteq \baseseven{\D1 (\Omega_7^* \setminus \D0\Omega_7^*)}$
is $\makeset{\baseseven{w}}{\baseseven{\D1w} \in S}$.
The value on $ S \subseteq \baseseven{\D1 \D0 \Omega_7^*}$ equals $\emptyset $.
\end{ourlemma}

With Lemma~\ref{removing_leading_digit_1_lemma} established
and the expression (\ref{removing_leading_digit_1_lemma__expression})
proved to implement the function $\removeone(X)$,
the notation $\removeone(X)$ 
is used in equations
to refer to this subexpression.

Next, consider the function $E(X)$
representing the existential quantifier
ranging over strings in $\{\D3, \D6\}^*$.
This function can be implemented by a single expression as follows:

\begin{lemmaE}[Representing the existential quantifier]\hacklabel{E}
\label{existential_quantifier_lemma}
The value of the expression 
\begin{equation*}
	\left( X \cap \baseseven{\D1 \Omega_7^*}\right)
	\cup 
	\left(
	\big[
	(X \cap \baseseven{\{\D3,\D6\}^+\D1 \Omega_7^*})
	\dotminus \baseseven{\{\D3,\D6\}^+\D0^*}
	\big]
	\cap \baseseven{\D1 \Omega_7^*}\right)
\end{equation*}
on any $S \subseteq \baseseven{\{\D3, \D6\}^*\D1\Omega_7^*}$
is $E(S)=\makeset{\baseseven{\D1w}}{\exists w' \in \{\D3, \D6 \}^* \baseseven{w'\D1w} \in S}$.
\end{lemmaE}

Note that $E(X)$ can already produce any recursively enumerable set
from a recursive argument,
and therefore it is essential to use subtraction in the expression.

With the existential quantifier implemented,
the next task is to represent a universal quantifier.
Ideally, one would be looking for an expression implementing $A(X)$,
but, unfortunately, no such expression was found,
and the actual construction given below
implements the universal quantifier
using multiple equations.
The first step is devising an equation
representing the function
$f(X)=\makeset{\baseseven{x\D1w}}{x \in \{\D3,\D6\}^*, \; \baseseven{\D1w} \in X}$,
which appends every string of digits in $\{\D3,\D6\}^*$
to numbers in its argument set.

\begin{ourlemma}\label{concatenating_36star_to_1w_lemma}
For every constant set $X \subseteq \baseseven{\D1 \Omega_7^*}$,
the equation
\begin{align*} 
Y &= X \cup \appendthreesix(Y), \quad \text{where} \\
\appendthreesix(Y)
	&=
\bigcup_{i,j \in \{\D3,\D6\}}
\Big[
\Big(
\big[
\big(Y \cap \baseseven{j \Omega_7^*} \big)
+ \baseseven{\D2\D0^*}\big]
\cap \baseseven{\D2j\Omega_7^*}
\Big)
+ \baseseven{(i-2)\D0^*}
\Big]
\cap \baseseven{ij\Omega_7^*} \\
&\cup
\bigcup_{i \in \{\D3,\D6\}}
\big[(Y \cap \baseseven{\D1 \Omega_7^*} )
+ \baseseven{i\D0^*}\big]
\cap \baseseven{i\D1\Omega_7^*}
\end{align*}
has the unique solution
$Y = \makeset{\baseseven{x\D1w}}{x \in \{\D3,\D6\}^*, \baseseven{\D1w} \in X}$.
\end{ourlemma}

\begin{lemmaA}[Representing the universal quantifier]\hacklabel{A}
\label{universal_quantifier_lemma}
Let $S,\widetilde{S} \subseteq \baseseven{\{\D3, \D6\}^*\D1\Omega_7^*}$
be any sets, such that $\widetilde{S} \cap S = \emptyset$ and
for $x',x \in \{\D3,\D6\}^*$
$\baseseven{x \D1 w} \in S$ and $\baseseven{x' \D1 w} \notin S$ implies
$\baseseven{x' \D1 w} \in \widetilde{S}$.  
Then the following system of equations over sets of integers
in variables $Y$, $\widetilde{Y}$ and $Z$ 
\begin{align*}
Y &= Z \cup \appendthreesix(Y) \\
\widetilde{Y} &= E(\widetilde{S}) \cup \appendthreesix(\widetilde{Y}) \\
Z &\subseteq \baseseven{\D1\Omega_7^+} \\
Y \subseteq 
	S &\subseteq Y \cup \widetilde{Y},
\end{align*}
has the unique solution
$Z = A(S) = \makeset{\baseseven{\D1w}}{\forall x \in \{\D3, \D6\}^*: \:
	\baseseven{x\D1w} \in S}$,
$Y = \makeset{\baseseven{y\D1w}}{y \in \{\D3, \D6\}^*,
	\forall x \in \{\D3, \D6\}^*: \: \baseseven{x\D1w} \in S}$,
$\widetilde{Y} = \makeset{\baseseven{y\D1w}}{y \in \{\D3, \D6\}^*, \:
	\exists x \in \{\D3, \D6\}^*: \:\baseseven{x\D1w} \in \widetilde{S}}$.
\end{lemmaA}

Once the above quantifiers
process a number $\baseseven{\D1 x_k \D1 x_{k-1} \ldots \D1 x_1 \D1 w}$,
reducing it to $\baseseven{\D1 w}$,
the actual number $\baseseven{w}$
is obtained from this encoding
by Lemma~\ref{removing_leading_digit_1_lemma}.

\begin{ourtheorem}\label{AH_theorem}
Every arithmetical set $S \subseteq \mathbb{Z}$
($S \subseteq \mathbb{N}$)
is representable as a component of a unique solution
of a system of equations
over sets of integers (sets of natural numbers, respectively)
with $\varphi_j, \psi_j$ using the operations
of addition and union and ultimately periodic constants
(addition, subtraction, union and singleton constants, respectively).
\end{ourtheorem}

\section{Representing hyper-arithmetical sets}\label{section_HA}

Following
Moschovakis~\citex{Moschovakis}{Sec.\,8E}
and Aczel~\citex{handbook}{Thm.\,2.2.3},
\emph{hyper-arithmetical} sets $B_1,B_2, \ldots$
shall be defined as the \emph{smallest effective $\sigma$-ring},
which is the recursion-theoretic counterpart to Borel sets
(the smallest family of sets containing all open sets and closed under
countable union and countable intersection).

Let $f_1, f_2 , \ldots$ be an enumeration of all partial recursive functions
and let $\tau_1$, $\tau_2$ be two recursive functions.
Then, for all $k \in \mathbb N$,
\begin{equation*}
	B_{\tau_1(k)} = \mathbb{N} \setminus \{k\}, \quad
	C_{\tau_1(k)} = \{k\} 
\end{equation*}
Moreover, for all numbers $k \in \mathbb{N}$,
if $f_k$ is a total function, then
\begin{equation*}
B_{\tau_2(k)} = \bigcup_{n \in \mathbb{N}} C_{f_k(n)}, \quad
C_{\tau_2(k)} = \bigcap_{n \in \mathbb{N}} B_{f_k(n)},
\end{equation*}
where the former operation is known as \emph{effective $\sigma$-union},
while the latter is \emph{effective $\sigma$-intersection}.
Note that the only distinction between $B_e$ and $C_e$
is that the former is defined as a union and the latter as an intersection.
As the definitions are dual, $B_e=\overline{C_e}$.

The family of sets $\mathcal B = \makeset{B_e,C_e}{e \in I}$,
where $I \subseteq \mathbb{N}$ is an index set, 
is called an effective $\sigma$-ring, if it contains $\makeset{B_{\tau_1(e)},C_{\tau_1(e)}}{e \in \mathbb{N}}$
and is closed under effective $\sigma$-union and effective $\sigma$-intersection.
Then the hyper-arithmetical sets
are defined as the smallest effective $\sigma$-ring,
which can be formally defined
as the least fixed point of a certain operator
on the set $\mathcal{A}=2^{\mathbb{N} \times 2^{\mathbb{N}} \times 2^{\mathbb{N}}}$,
where a triple $(e, B_e, C_e)$
indicates that the sets $B_e$ and $C_e$
have been defined for the index $e$ in the above inductive definition,
and an operator $\Phi : \mathcal{A} \to \mathcal{A}$
represents one step of this inductive definition.
Furthermore, this least fixed point
can be obtained constructively
by a transfinite induction on countable ordinals,
which is essential for any proofs about hyper-arithmetical sets.
It is known \citex{Moschovakis}{Sec. 8E} \citex{handbook}{Thm. 2.2.3}
that for some (easy) choices of $\tau_1$ and $\tau_2$ the smallest
effective $\sigma$-ring coincides with $\Delta_1^1$ sets.
Fix those two functions and the corresponding $\mathcal B$.
Note that the definition is valid
not for every choice of $\tau_1$ and $\tau_2$:
in particular, they must be one-to-one and have disjoint images.

With every set $B_e \in \mathcal B$
one can associate a \emph{tree of $B_e$},
labelled with
sets from $\mathcal B$:
its root is labelled with $B_e$,
and each vertex $B_{\tau_2(e')}$
($C_{\tau_2(e')}$, respectively)
in the tree has children
labelled with $\makeset{C_{f_{e'}(n)}}{n \in \mathbb N}$
($\makeset{B_{f_{e'}(n)}}{n \in \mathbb N}$, respectively).
Vertices of the form $B_{\tau_1(e')}$ or $C_{\tau_1(e')}$
have no children; these are the only leaves in the tree.

A partial order $\prec$ is \emph{well-founded},
if it has no infinite descending chain.
Extending this notion to oriented trees, a tree is well-founded
if it contains no infinite downward path.

\begin{ourlemma}
\label{lemma-trees-are-well-founded}
For each pair of sets $B_e,C_e \in \mathcal B$
the trees of $B_e,C_e$ are well-founded.
\end{ourlemma}

The well-foundedness of a set allows using the \emph{well-founded induction principle}:
given a property $\phi$ and a well founded order $\prec$ on a set $A$,
$\phi(n)$ is true for all $n \in A$
if
$$
(\forall m \prec n \; \phi(m) ) \Rightarrow \phi(n).
$$
This principle shall be used in the proof of the main construction,
which is described in the rest of this section.
Note, that the basis of the induction
are $\prec$-minimal elements $n$ of $A$,
as for them $\phi(n)$ has to be shown directly.

Fix $B_{i_0}$ as the target set in the root.
Consider a path of length $k$ in this tree,
going from $B_{i_0}$ to $C_{i_1}$, $B_{i_2}$, \ldots, $B_{i_k}$
(or $C_{i_k}$, depending on the parity of $k$).
Then, for each $j$-th set in this path,
$i_j = f_{\tau_2^{-1}(i_{j-1})}(n_j)$ for some number $n_j$,
and the path is uniquely defined by the sequence of numbers $n_1, \ldots, n_k$.
Consider the binary encoding of each of these numbers
written using digits $\D3$ and $\D6$ (representing zero and one, respectively),
and let $\Resolve$ 
be a partial function that maps
finite sequences of such ``binary'' strings
representing numbers $n_1, \ldots, n_k$
to the number $i_k$ of the set $B_{i_k}$ or $C_{i_k}$ in the end of this path.
The value of this function can be formally defined by induction:
\begin{align*}
	\Resolve(\langle\rangle) = i_0, \qquad
	\Resolve(x_1, \ldots, x_k) =
	f_{\tau_2^{-1}(\Resolve(x_1, \ldots, x_{k-1}))}(\basetwo{x_k}),
\end{align*}
Note that $\Resolve$ may be undefined
if some $\tau_2$-preimage is undefined.

The goal is to construct a system of equations,
such that the following two sets are
among the components of its unique solution:
\begin{align*}
	\Goal_0
		&=
	\makeset{\baseseven{\D1 x_k \D1 x_{k-1} \ldots \D1 x_1 \D1\D0 w}}{
	k \geqslant 0,
	x_i \in \{\D3, \D6\}^*,
	\baseseven{w} \in B_{\Resolve(x_1, \ldots, x_k)}
	}, \\
	\Goal_1
		&=
	\makeset{\baseseven{\D1 x_k \D1 x_{k-1} \ldots \D1 x_1 \D1\D0 w}}{
	k \geqslant 0,
	x_i \in \{\D3, \D6\}^*,
	\baseseven{w} \in C_{\Resolve(x_1, \ldots, x_k)}
	}.
\end{align*}
These sets encode the sets $B_0, B_1, \ldots $ needed to compute $B_{i_0}$.
In this way the (possibly infinite) amount of equations defining sets
in hyper-arithmetical hierarchy is encoded in a finite amount of equations
using only small number of variables.
The set $B_i$ in the node with path to the root encoded by $x_k,x_{k-1}, \ldots, x_1 \in \{ \D3, \D6\}^*$
is represented by
$\makeset{\baseseven{\D1 x_k \D1 \ldots  \D1 x_k \D1 \D0 w}}{\baseseven{w} \in B_i}
\subseteq
\Goal_0$.

The following set defines the admissible encodings,
that is, numbers encoding paths in the tree of $B_{i_0}$:
\begin{equation*}
	\Admissible
		=
	\makeset{\baseseven{\D1 x_k \D1 x_{k-1} \D1 \ldots \D1 x_1 \D1\D0 w}}{%
	k \geqslant 0, \:
	x_i \in \{\D3, \D6\}^*, \:
	\Resolve(x_1, \ldots, x_k)
	\text{\ is defined}
	}
\end{equation*}
The next two sets represent
the leaves of the tree of $B_{i_0}$,
and the numbers in those leaves:
\begin{multline*}
	R_0
		=
	\{\baseseven{\D1 x_k \D1 x_{k-1} \ldots \D1 x_1 \D1\D0 w} \: | \\
	k \geqslant 0, \:
	x_i \in \{\D3, \D6\}^*, \exists e \in \mathbb{N}:
	\Resolve(x_1, \ldots, x_k) = \tau_1(e), \:
	\baseseven{w} \in B_{\tau_1(e) }\}, \\
\end{multline*}
\begin{multline*}
	R_1
		=
	\{\baseseven{\D1 x_k \D1 x_{k-1} \ldots \D1 x_1 \D1\D0 w} \: | \\
	k \geqslant 0, \:
	x_i \in \{\D3, \D6\}^*, \exists e \in \mathbb{N}:
	\Resolve(x_1, \ldots, x_k) = \tau_1(e), \:
	\baseseven{w} \in C_{\tau_1(e)}\}.
\end{multline*}

\begin{ourlemma}\label{constants_are_recursive}
The sets $\Goal_i$, $\Admissible$, $R_i$ are r.e.\ sets,
$\Resolve$ is an r.e.\ predicate.
\end{ourlemma}

Consider the following system of equations:
\begin{align}
\label{eq:main-X0}
X_0 &= E(\removeone(X_1)) \cup R_0 \\
\label{eq:main-X1}
X_1 &= Z \cup R_1 \\
\label{eq:main-widetildeY}
\widetilde{Y} &=  E(\removeone(X_1)) \cup \appendthreesix(\widetilde{Y})\\
\label{eq:main-Y}
Y &= Z \cup \appendthreesix(Y) \\
\label{eq:main-inclusions}
Y &\subseteq \removeone(X_0 \cap \Admissible ) \subseteq Y \cup \widetilde{Y} \\
\label{eq:main-Z}
Z &\subseteq \baseseven{\D1\Omega_7^+} \\
\label{eq:main-X0X1-admissible}
X_0,X_1 &\subseteq \Admissible \\
\label{eq:main-X0-R1-intersection-is-empty}
X_0 \cap R_1 &= X_1 \cap R_0 = \emptyset
\end{align}

Its intended unique solution
has $X_0=\Goal_0$ and $X_1=\Goal_1$,
and
accordingly encodes the set $B_{i_0}$,
as well as all sets of $\mathcal B$
on which $B_{i_0}$ logically depends.
The system implements the functions $E(X)$ and $A(X)$
to represent effective $\sigma$-union and $\sigma$-intersection,
respectively.
For that purpose,
the expression for $E(X)$ introduced in Lemma~\ref{existential_quantifier_lemma},
as well as the system of equations implementing $A(X)$
defined in Lemma~\ref{universal_quantifier_lemma},
are applied iteratively to the same variables $X_0$ and $X_1$.
Intuitively,
the above system may be regarded as an implementation
of an equation $X_0=A(E(X_0)) \cup const$.

The proof uses the principle of induction on well-founded structures.
The membership of numbers of the form
$\baseseven{\D1 x_k \D1 x_{k-1} \ldots \D1 x_1 \D1\D0 w}$
in the variables $X_0$ and $X_1$,
where $k \geqslant 0$,
$x_i \in \{\D3, \D6\}^*$
and $w \in \Omega_7^* \setminus \D0 \Omega_7^*$,
is first proved for larger $k$'s
and then inductively extended down to $k=0$,
which allows extracting $B_{i_0}$ out of the solution.
The well-foundedness of the tree of $B_{i_0}$
means that although $B_{i_0}$ depends upon infinitely many sets,
each dependency is over a finite path ending with a constant,
that is, the self-dependence of numbers in $X_0,X_1$
on the numbers in $X_0,X_1$
reaches a constant $R_0,R_1$ in finitely many steps
(yet the number of steps is unbounded).

\begin{ourlemma}
\label{goal-is-unique-solution-lemma}
The unique solution of the system
\eqref{eq:main-X0}--\eqref{eq:main-X0-R1-intersection-is-empty}
is
\begin{align*}
X_0 &= \Goal_0
		=
	\makeset{\baseseven{\D1 x_k \ldots \D1 x_1 \D1\D0 w}}{%
	k \geqslant 0,
	x_i \in \{\D3, \D6\}^*,
	\baseseven{w} \in B_{\Resolve(x_1, \ldots, x_k)}
	}\\
X_1 &= \Goal_1
		=
	\makeset{\baseseven{\D1 x_k \ldots \D1 x_1 \D1\D0 w}}{%
	k \geqslant 0,
	x_i \in \{\D3, \D6\}^*,
	\baseseven{w} \in C_{\Resolve(x_1, \ldots, x_k)}
	}\\
Y &= 	\makeset{\baseseven{x_{k+1} \D1 x_k \ldots \D1 x_1 \D1\D0 w}}{%
	k \geqslant 0,
	x_i \in \{\D3, \D6\}^*,
	\forall x_{k+1}:
	\baseseven{w} \in B_{\Resolve(x_1, \ldots, x_{k+1})}
	}\\
\widetilde{Y} &= 
	\makeset{\baseseven{x_{k+1} \D1 x_k \ldots \D1 x_1 \D1\D0 w}}{%
	k \geqslant 0,
	x_i \in \{\D3, \D6\}^*,
	\exists x_{k+1}:
	\baseseven{w} \in C_{\Resolve(x_1, \ldots, x_{k+1})}
	}\\
Z &= \Goal_1 \setminus R_1
		=
	\{\baseseven{\D1 x_k \ldots \D1 x_1 \D1\D0 w} \: \mid \\
	& \hspace*{3cm}
	k \geqslant 0, e\in \mathbb{N}, 
	x_i \in \{\D3, \D6\}^*,
	\Resolve(x_1, \ldots, x_k) = \tau_2(e),
	\baseseven{w} \in C_{\tau_2(e)}
	\}
\end{align*}
\end{ourlemma}

Then, in order to obtain the set $B_{i_0}$,
it remains to intersect $X_0=\Goal_0$
with the recursive constant set $\baseseven{\D{10}\Omega_7^*}$,
and then remove the leading digits $\D{10}$
by a construction analogous 
to the one in Lemma~\ref{removing_leading_digit_1_lemma}.

\begin{ourtheorem}\label{HA_theorem}
For every hyper-arithmetical set $B \subseteq \mathbb{Z}$
($B \subseteq \mathbb{N}$)
there is a system of equations over subsets of $\mathbb{Z}$
(over subsets of $\mathbb{N}$, respectively)
using union, addition and ultimately periodic constants
(union, addition, subtraction and singleton constants, respectively),
such that $(B, \ldots)$ is its unique solution.
\end{ourtheorem}

\section{Equations with addition only}\label{section_addition_only}

Equations over sets of natural numbers
with addition as the only operation
can represent an \emph{encoding} of every recursive set,
with each number $n \in \mathbb{N}$
represented by the number $16n+13$ in the encoding
\cite{EquationsUnaryPlus}.
In order to define this encoding,
for each $i \in \{0, 1, \ldots, 15\}$
and for every set $S \subseteq \mathbb{Z}$, denote:
\begin{align*}
	\embed{i}{S} = \makeset{16n+i}{n \in S}.
\end{align*}
The encoding of a set of natural numbers $\widehat{S} \subseteq \mathbb{N}$
is defined as
\begin{equation*}
	S=\sigma_0(\widehat{S})
		=
	\{0\}
	\cup \embed{6}{\mathbb{N}}
	\cup \embed{8}{\mathbb{N}}
	\cup \embed{9}{\mathbb{N}}
	\cup \embed{12}{\mathbb{N}}
	\cup \embed{13}{\widehat{S}},
\end{equation*}

\begin{ourproposition}[\citex{EquationsUnaryPlus}{Thm.~5.3}]
For every recursive set $S$
there exists a system of equations over sets of natural numbers
in variables $X, Y_1, \ldots, Y_m$
using the operation of addition and ultimately periodic constants,
which has a unique solution with $X=\sigma_0(S)$.
\end{ourproposition}
This result is proved by first representing the set $S$
by a system with addition and union,
and then by representing addition and union of sets
using addition of their $\sigma_0$-encodings.

The purpose of this section is to obtain a similar result
for equations over sets of integers:
namely, that they can represent the same kind of encoding
of every hyper-arithmetical set.
For every set $\widehat{S} \subseteq \mathbb{Z}$,
define its \emph{encoding} as the set
\begin{equation*}
	S=\sigma(\widehat{S})
		=
	\{0\}
	\cup \fulltrack{6}
	\cup \fulltrack{8}
	\cup \fulltrack{9}
	\cup \fulltrack{12}
	\cup \embed{13}{\widehat{S}}.
\end{equation*}
The subset $S \cap \makeset{16n+i}{n \in \mathbb{Z}}$
is called the \emph{$i$-th track of $S$}.

The first result on this encoding
is that the condition of a set $X$
being an encoding of any set
can be specified by an equation of the form $X+C=D$.

\begin{ourlemma}[cf. \citex{EquationsUnaryPlus}{Lemma~3.3}]
\label{good_encoding_lemma}
A set $X \subseteq \mathbb{Z}$ satisfies an equation
\begin{equation*}
X + \{0, 4, 11\}
=
\bigcup_{\substack{i \in \{ 0,1,3,4,6,7, \\ 8,9,10,12,13\}}} \embed{i}{\mathbb{Z}}
\cup \{ 11 \}
\end{equation*}
if and only if 
$X = \sigma(\widehat{X})$ for some $\widehat{X} \subseteq \mathbb{Z}$.
\end{ourlemma}

Now, assuming that
the given system of equations with union and addition
is decomposed to have all equations of the form
$X=Y+Z$, $X=Y \cup Z$ or $X=const$,
these equations can be simulated in a new system as follows:

\begin{ourlemma}[cf. \citex{EquationsUnaryPlus}{Lemma~4.1}]
\label{transforming_single_equation_lemma}
For all sets $X, Y, Z \subseteq \mathbb{Z}$,
\begin{align*}
	\sigma(Y) + \sigma(Z) + \{0, 1\} &= \sigma(X) + \sigma(\{0\}) + \{0, 1\}
		\quad\text{if and only if}\quad
	Y + Z = X \\
	\sigma(Y) + \sigma(Z) + \{0, 2\} &= \sigma(X) + \sigma(X) + \{0, 2\}
		\quad\text{if and only if}\quad
	Y \cup Z = X.
\end{align*}
\end{ourlemma}

Using these two lemmata,
one can simulate any system with addition and union
by a system with addition only.
Taking systems representing different hyper-arithmetical sets,
the following result on the expressive power of systems with addition
can be established:

\begin{ourtheorem}\label{only_addition_theorem}
For every hyper-arithmetical set $S \subseteq \mathbb{Z}$
there exists a system of equations
over sets of integers
using the operation of addition
and ultimately periodic constants,
which has a unique solution with $X_1=T$,
where $S=\makeset{n}{16n \in T}$.
\end{ourtheorem}

\section{Decision problems} \label{section_decision}

Having a solution (solution existence)
and having exactly one solution (solution uniqueness)
are basic properties of a system of equations.
For language equations with continuous operations,
\emph{solution existence} is $\Pi^0_1$-complete~\cite{DecisionProblems},
and it remains $\Pi^0_1$-complete already in the case of a unary alphabet,
concatenation as the only operation and regular constants~\cite{EquationsUnaryPlus},
that is, for equations over sets of natural numbers with addition only.
For the same formalisms,
\emph{solution uniqueness} is $\Pi^0_2$-complete.

Consider equations over sets of integers.
Since their expressive power extends beyond the arithmetical hierarchy,
the decision problems should accordingly be harder.
In fact, the solution existence is $\Sigma^1_1$-complete,
which will now be proved
using a reduction from the following problem:

\begin{ourproposition}[Rogers~\citex{Rogers}{Thm.~16-XX}]
\label{pi_1_1_complete_problem_on_trees_proposition}
Consider trees with nodes labelled by finite sequences of natural numbers,
such that a node $(x_1, \ldots, x_{k-1}, x_k)$
is a son of $(x_1, \ldots, x_{k-1})$,
and the empty sequence $\epsilon$ is the root.
Then the following problem is $\Pi^1_1$-complete:
``Given a description of a Turing machine
recognizing the set of nodes of a certain tree,
determine whether this tree has no infinite paths''.
\end{ourproposition}

In other words, a given Turing machine
recognizes sequences of natural numbers,
and the task is to determine whether
there is \emph{no} infinite sequence of natural numbers,
such that all of its prefixes are accepted by the machine.
The $\Sigma^1_1$-complete complement of the problem
is testing whether such an infinite sequence exists,
and it can be reformulated as follows:

\begin{ourcorollary}\label{sigma_1_1_complete_problem_on_trees_corollary}
The following problem is $\Sigma^1_1$-complete:
``Given a Turing machine $M$ working on natural numbers,
determine whether there exists
an infinite sequence of strings $\{x_i\}_{i=1}^\infty$
with $x_i \in \{\D3, \D6\}^*$,
such that $M$ accepts
$\baseseven{\D1 x_k \D1 x_{k-1} \ldots \D1 x_1 \D1}$
for all $k \geqslant 0$''.
\end{ourcorollary}
This problem can be reduced
to testing existence of a solution of equations over sets of numbers.

\begin{ourtheorem}\label{solution_existence_sigma_1_1_theorem}
The problem of whether a given system of equations
over sets of integers
with addition and ultimately periodic constants
has a solution
is $\Sigma^1_1$-complete.
\end{ourtheorem}

\begin{table}
\begin{tabular}{|l|l|l|l|}
\hline
	& Sets representable
		& \multicolumn{2}{|c|}{Complexity of decision problems}
			\\
\cline{3-4}
	& by unique solutions
		& solution existence
			& solution uniqueness
				\\
\hline
over $2^\mathbb{N}$, with $\{+, \cup\}$
	& $\Delta^0_1$ (recursive) \cite{EquationsUnary}
		& $\Pi^0_1$-complete \cite{EquationsUnary}
			& $\Pi^0_2$-complete \cite{EquationsUnary}
				\\
over $2^\mathbb{N}$, with $\{+\}$
	& encodings of $\Delta^0_1$ \cite{EquationsUnaryPlus}
		& $\Pi^0_1$-complete \cite{EquationsUnaryPlus}
			& $\Pi^0_2$-complete \cite{EquationsUnaryPlus}
				\\
\hline
over $2^\mathbb{N}$, with $\{+, \dotminus, \cup\}$
	& $\Delta^1_1$ (hyper-arithmetical)
		& $\Sigma^1_1$-complete
			& $\Pi^1_1 \leqslant \cdot \leqslant \Delta^1_2$
				\\
over $2^\mathbb{Z}$, with $\{+, \cup\}$
	& $\Delta^1_1$
		& $\Sigma^1_1$-complete
			& $\Pi^1_1 \leqslant \cdot \leqslant \Delta^1_2$
				\\
over $2^\mathbb{Z}$, with $\{+\}$
	& encodings of $\Delta^1_1$
		& $\Sigma^1_1$-complete
			& $\Pi^1_1 \leqslant \cdot \leqslant \Delta^1_2$
				\\
\hline
\end{tabular}
\caption{Summary of the results.}
\label{t:summary}
\end{table}

Now consider the solution uniqueness property.
The following upper bound on its complexity
naturally follows by definition:

\begin{ourtheorem}\label{solution_uniqueness_delta_1_2_proposition}
The problem of whether a given system of equations over sets of integers
using addition and ultimately periodic constants
has a unique solution
can be represented as a conjunction
of a $\Sigma^1_1$-formula
and a $\Pi^1_1$-formula,
and is accordingly in $\Delta^1_2$.
At the same time, the problem is $\Pi^1_1$-hard.
\end{ourtheorem}

The exact hardness of testing solution uniqueness is still open.
The properties of different families of equations over sets of numbers
are summarized in Table~\ref{t:summary}.


\begin{thebibliography}{99}
\bibitem{handbook} P. Aczel,
	{``An introduction to inductive definitions''},
	in: J. Barwise (Ed.),
	\emph{Handbook of Mathematical Logic},
	739--783,
	North-Holland, 1977.
\bibitem{dAlessandroSakarovitch} F. d'Alessandro, J. Sakarovitch,
	{``The finite power property in free groups''},
	\emph{Theoretical Computer Science},
	293:1 (2003), 55--82.
\bibitem{Anisimov} A. V. Anisimov,
	{``Languages over free groups''},
	\emph{Mathematical Foundations of Computer Science},
	(MFCS 1975, Mari\'ansk\'e L\'azn\v{e}, September 1--5, 1975),
	LNCS 32, 167--171.
\bibitem{GinsburgRice} S. Ginsburg, H. Rice,
	{``Two families of languages related to ALGOL''},
	\emph{J. of the ACM},
	9 (1962), 350--371.
\bibitem{Halpern} J. Y. Halpern,
	{``Presburger arithmetic with unary predicates is $\Pi^1_1$ complete''},
	\emph{Journal of Symbolic Logic},
	56:2 (1991), 637--642.
\bibitem{Jez_DLT} A. Je\.z,
	{``Conjunctive grammars can generate non-regular unary languages''},
	\emph{International Journal of Foundations of Computer Science},
	19:3 (2008), 597--615.
\bibitem{ConjunctiveUnary} A. Je\.z, A. Okhotin,
	{``Conjunctive grammars over a unary alphabet: undecidability and unbounded growth''},
	\emph{Theory of Computing Systems},
	46:1 (2010), 27--58.
\bibitem{EquationsUnary} A. Je\.z, A. Okhotin,
	{``On the computational completeness of equations over sets of natural numbers''}
	\emph{ICALP 2008}
	(Reykjavik, Iceland, July 7--11, 2008),
	LNCS 5126, 63--74.
\bibitem{EquationsUnaryPlus} A. Je\.z, A. Okhotin,
	{``Equations over sets of natural numbers with addition only''},
	\emph{STACS 2009}
	(Freiburg, Germany, 26--28 February, 2009),
	577--588.
\bibitem{Kunc} M. Kunc,
	{``The power of commuting with finite sets of words''},
	\emph{Theory of Computing Systems},
	40:4 (2007), 521--551.
\bibitem{Kunc_survey} M. Kunc,
	{``What do we know about language equations?''},
	\emph{Developments in Language Theory}
	(DLT 2007, Turku, Finland, July 3--6, 2007),
	LNCS 4588, 23--27.
\bibitem{EquationsUnaryPlus2} T. Lehtinen, A. Okhotin,
	{``On equations over sets of numbers and their limitations''},
	\emph{Developments in Language Theory}
	(DLT 2009, Stuttgart, Germany, 30 June--3 July, 2009),
	LNCS 5583, 360--371.
\bibitem{McKenzieWagner} P. McKenzie, K. Wagner,
	{``The complexity of membership problems
	for circuits over sets of natural numbers''},
	\emph{Computational Complexity},
	16:3 (2007), 211--244.
\bibitem{Moschovakis} Y. Moschovakis,
	\emph{Elementary Induction on Abstract Structures},
	North-Holland, 1974.
\bibitem{Conjunctive} A. Okhotin,
	``Conjunctive grammars'',
	\emph{Journal of Automata, Languages and Combinatorics},
	6:4 (2001), 519--535.
\bibitem{ConjEquations} A. Okhotin,
	{``Conjunctive grammars and systems of language equations''},
	\emph{Programming and Computer Software},
	28:5 (2002), 243--249.
\bibitem{EquationsUnresolved} A. Okhotin,
	{``Unresolved systems of language equations: expressive power and decision problems''},
	\emph{Theoretical Computer Science},
	349:3 (2005), 283--308.
\bibitem{EquationsUniversality} A. Okhotin,
	``Computational universality in one-variable language equations'',
	\emph{Fundamenta Informaticae},
	74:4 (2006), 563--578.
\bibitem{DecisionProblems} A. Okhotin,
	{``Decision problems for language equations''},
	\emph{Journal of Computer and System Sciences},
	76 (2010), to appear;
	earlier version at ICALP 2003.
\bibitem{Robinson} J. Robinson,
	{``An introduction to hyperarithmetical functions''},
	\emph{Journal of Symbolic Logic},
	32:3 (1967), 325--342.
\bibitem{Rogers} H. Rogers, Jr.,
	\emph{Theory of Recursive Functions and Effective Computability},
	McGraw-Hill, 1967.
\bibitem{Travers} S. D. Travers,
	{``The complexity of membership problems for circuits over sets of integers''}
	\emph{Theoretical Computer Science},
	369:1--3 (2006), 211--229.
\end{thebibliography}
\end{document}